\documentclass[10pt]{article}
\usepackage{latexsym,graphicx}
\newcommand{\be}{\begin{equation}}
\newcommand{\ee}{\end{equation}}
\def\n{\noindent}
\catcode `\@=11
 \catcode `\@=12
\begin{document}
\begin{center}
\large{\bf{Bianchi Type-I Anisotropic Dark Energy Models with Constant Deceleration Parameter}} \\
\vspace{10mm}
\normalsize{Anirudh Pradhan$^1$, H. Amirhashchi $^2$, Bijan Saha$^3$}\\
\vspace{5mm}
\normalsize{$^{1}$Department of Mathematics, Hindu Post-graduate College, Zamania-232 331, 
Ghazipur, India \\
E-mail: pradhan@iucaa.ernet.in} \\
\vspace{5mm}
\normalsize{$^{2}$Department of Physics, Mahshahr Islamic Azad University, Mahshahr, Iran \\ 
E-mail: hashchi@yahoo.com} \\
\vspace{5mm}
\normalsize{$^{1,3}$Laboratory of Information Technologies, Joint Institute for Nuclear Research, 141880 Dubna, Russia \\
E-mail: bijan@jinr.ru } \\
\end{center}
\vspace{10mm}
\begin{abstract} 
New dark energy models in anisotropic Bianchi type-I (B-I) space-time with 
variable EoS parameter and constant deceleration parameter have been investigated in the present paper. The 
Einstein's field equations have been solved by applying a variation law for generalized Hubble's parameter in 
B-I space-time. The variation law for Hubble's parameter generates two types of solutions for the average 
scale factor, one is of power-law type and other is of the exponential form. Using these two forms, Einstein's 
field equations are solved separately that correspond to expanding singular and non-singular models of the universe 
respectively. The equation of state (EoS) parameter $\omega$ is found to be time dependent and its existing range 
for this model is in good agreement with the recent observations of SNe Ia data, SNe Ia data (with CMBR anisotropy) 
and galaxy clustering statistics. The cosmological constant $\Lambda$ is found to be a decreasing function of time 
and it approaches a small positive value at late time (i.e. the present epoch) which is corroborated by results 
from recent supernovae Ia observations. 
\end{abstract}
 \smallskip
\n Keywords : Bianchi type-I universe, Dark energy, Variable EoS parameter \\
\n PACS number: 98.80.Es, 95.36.+x 
\section{Introduction}
Before few years ago two groups (the Supernova Cosmology Project and the High-Z Supernova Team) presented 
evidence that the expansion of the universe is accelerating (Garnavich et al. \cite{ref1,ref2}; 
Perlmutter et al. \cite{ref3,ref4,ref5}; Riess et al. \cite{ref6}; Schmidt et al. \cite{ref7}. These teams 
have measured the distances to cosmological supernovae by using the fact that the intrinsic luminosity of 
Type Ia supernovae is closely correlated to their decline rate from maximum brightness, which can be 
independently measured. These measurements, combined with red-shift data for the supernovae, led to the 
prediction of an accelerating universe. Both team obtained $\Omega_{M} \approx 0.3$, $\Omega_{\Lambda} \approx 0.7$, 
and strongly ruled out the traditional ($\Omega_{M}, \Omega_{\Lambda}$) = ($1, 0$) universe. This value of the 
density parameter $\Omega_{\Lambda}$ corresponds to a cosmological constant that is small, nevertheless, nonzero 
and positive, that is, $\Lambda \approx 10^{-52}m^{-2} \approx 10^{-35} s^{-2}$. An intense search is going on, in 
both theory and observations, to unveil the true nature of this acceleration. It is commonly believed by the 
cosmological community that a kind of repulsive force which acts as anti-gravity is responsible for gearing up 
the Universe some $7$ billion years ago. This hitherto unknown exotic physical entity is termed as {\it dark energy}. \\

It is an irony of nature and is a puzzling phenomenon that most abundant form of matter-energy in the universe 
is most mysterious. The simplest Dark Energy (DE) candidate is the cosmological constant $\Lambda$, but it needs to 
be extremely fine-tuned to satisfy the current value of the DE. Alternatively, to explain decay of
the density, many dynamic models have been suggested, where $\Lambda$ varies slowly with cosmic time (t)
(Overduin and Cooperstock \cite{ref8}; Sahni and Starobinsky \cite{ref9}; Komatsu et al. \cite{ref10}). In addition 
to models with dynamic $\Lambda(t)$, many hydrodynamic models with or without dissipative pressure have been proposed 
in which barotropic fluid is the source of DE (Sahni and Starobinsky \cite{ref9}; Komatsu et al. \cite{ref10}). 
Chaplygin gas as well as generalized Chaplygin gas have also been considered as possible dark energy sources due to 
negative pressure (Srivastava \cite{ref11}; Jackiw \cite{ref12}; Bertolami et al. \cite{ref13}; Bento, Bertolami and 
Sen \cite{ref14}; Bilic, Tupper and Viollier \cite{ref15}; Avelino et al. \cite{ref16}). Other than these approaches, 
some authors have considered modified gravitational action by adding a function $f(R)$ (R being the Ricci scalar 
curvature) to Einstein-Hilbert Lagrangian, where $f(R)$ provides a gravitational alternative for DE causing late-time 
acceleration of the universe (Capozziello\cite{ref17}; Caroll et al. \cite{ref18}; Dolgov and Kawasaki \cite{ref19}; 
Nojiri and Odintsov \cite{ref20,ref21}; Abdalaa et al. \cite{ref22}; Mena et al. \cite{ref23}). A review on modified 
gravity as an alternative to DE is available in Nojiri and Odintsov\cite{ref24}. A more comprehensive review is 
provided in Copeland et al. \cite{ref25}). Recently Gupta and Pradhan \cite{ref26} have presented an entirely new 
approach that cosmological nuclear energy is a possible candidate for DE. All these models are phenomenological in 
the sense that an idea of DE is introduced {\it a priori} either in term of gravitational field or non-gravitational 
field. In spite of these attempts, still cosmic acceleration is a challenge. \\

At present, there is great interest in cosmological models with a variable matter equation of state in the 
class of equation $p = \omega(t) \rho$ (p is the fluid pressure and $\rho$ its energy density). By now, methods 
allowing for restoration of the quantity $\omega(t)$ from expressional data have been developed (Sahni and 
Starobinsky \cite{ref27}, and an analysis of the experimental data has been conducted to determine this parameter 
as a function of cosmological time (see Sahni et al. \cite{ref28} and references therein). DE has been 
conventionally characterized by the equation of state (EoS) parameter mentioned above which is not necessarily 
constant. Recently, the parameter $\omega(t)$ is calculated with some reasoning which reduced to some simple 
parametrization of the dependences by some authors (Huterer and Turner\cite{ref29}; Weller and Albrecht \cite{ref30}; 
Chevallier and Polarski \cite{ref31}; Krauss et al. \cite{ref32}; Usmani et al. \cite{ref33}; Chen et al. 
\cite{ref34}. The simplest DE candidate is the vacuum energy ($\omega = - 1$), which is mathematically equivalent 
to the cosmological constant ($\Lambda$). The other conventional alternatives, which can be described by minimally 
coupled scalar fields, are quintessence ($\omega > - 1$), phantom energy $(\omega < - 1$ and quintom (that can 
across from phantom region to quintessence region as evolved) and have time dependent EoS parameter. Some other limits 
obtained from observational results coming from SN Ia data (Knop et al. \cite{ref35} collaborated with CMBR anisotropy 
and galaxy clustering statistics (Tegmark et al. \cite{ref36}) are $-1.67 < \omega < -0.62$ and $-1.33 < \omega 
< - 0.79$ respectively. However, it is not at all obligatory to use a constant value of $\omega$. Due to lack 
observational evidence in making a distinction between constant and variable $\omega$, usually the equation of state 
parameter is considered as a constant (Kujat et al. \cite{ref37}; Bartelmann et al. \cite{ref38}) with phase wise value 
$-1, 0, - \frac{1}{3}$ and $ + 1$ for vacuum fluid, dust fluid, radiation and stiff dominated universe, 
respectively. But in general, $\omega$ is a function of time or redshift (Jimenez \cite{ref39}; Das et al. 
\cite{ref40}; Ratra and Peebles \cite{ref41}). For instance, quintessence models involving scalar fields give rise 
to time dependent EoS parameter $\omega$ (Turner and White \cite{ref42}; Caldwell et al. \cite{ref43}; Liddle and 
Scherer \cite{ref44}; Steinhardt et al. \cite{ref45}). Some literature are also available on models with varying 
fields, such as cosmological models with variable equation of state parameter in Kaluza-Klein metric and wormholes 
(Rahaman et al. \cite{ref46,ref47}). In recent years various form of time dependent $\omega$ have been used for 
variable $\Lambda$ models (Mukhopadhyay et al. \cite{ref48}; Usmani et al. \cite{ref33}). Recently Ray et al. 
\cite{ref49}, Akarsu and Kilinc \cite{ref50} and Yadav \cite{ref51,ref52} have obtained dark energy models with 
variable EoS parameter. \\ 

Spatially homogeneous and anisotropic cosmological models play a significant role in the description of large 
scale behaviour of universe and such models have been widely studied in framework of General Relativity in search 
of a realistic picture of the universe in its early stages. Recently, Pradhan et al. \cite{ref53}$-$\cite{ref56} 
and Saha et al. \cite{ref57}$-$\cite{ref59} have studied homogeneous and anisotropic B-I space-time in different 
contexts. In this paper, we have investigated a new anisotropic B-I DE model with variable $\omega$ by assuming 
constant deceleration parameter. The out line of the paper is as follows: In Section 2, the metric and the field 
equations are described. Section 3 deals with the solutions of the field equations in two different cases with 
physical and geometric behaviour of the models. Finally, conclusions are summarized in the last Section 4.  

\section{The Metric and Field  Equations}
We consider totally anisotropic Bianchi type-I line element, given by
\begin{equation}
\label{eq1}
ds^{2} = - dt^{2}+ A^{2}dx^{2} + B^{2} dy^{2} + C^{2} dz^{2},
\end{equation}
where the metric potentials $A$, $B$ and $C$ are functions of $t$ alone. This ensures that the model is
spatially homogeneous. \\\\
The simplest generalization of Equation of State (EoS) parameter of perfect fluid may be to determine the EoS 
parameter separately on each spatial axis by preserving the diagonal form of the energy momentum tensor in a 
consistence way with the considered metric. Therefore, the energy momentum tensor of fluid is taken as
\begin{equation}
\label{eq2}
T_{ij} = diag [T_{00}, T_{11}, T_{22}, T_{33}]
\end{equation}
Thus, one may parameterize it as follows,
\[
T_{ij} = diag[\rho, - p_{x}, - p_{y}, - p_{z}] = diag[1, - \omega_{x}, - \omega_{y}, - \omega_{z}]\rho
\]
\begin{equation}
\label{eq3} =  diag[1, - \omega, - (\omega + \delta), - (\omega + \gamma)]\rho.
\end{equation}
Here $p_{x}, p_{y}$ and $p_{z}$ are the pressures, $\rho$ is the proper energy density
and $\omega_{x}, \omega_{y}$ and $\omega_{z}$ are the directional EoS parameters along the $x$, $y$ and $z$ axes 
respectively. $\omega$ is the deviation-free EoS parameter of the fluid. We have parameterized the deviation 
from isotropy by setting $\omega_{x}= \omega$ and then introducing skewness parameters $\delta$ and $\gamma$ 
that are the deviations from $\omega$ along the $y$ and $z$ axes respectively.\\\\
The Einstein's field equations (with gravitational units, $8\pi G = 1$ and $c = 1$) read as
\begin{equation}
\label{eq4} R_{ij} - \frac{1}{2} R g_{ij} = - T_{ij},
\end{equation}
where the symbols have their usual meaning. In a comoving co-ordinate system, Einstein's field equation (\ref{eq4}), 
with (\ref{eq3}) for B-I metric (\ref{eq1}) subsequently lead to the following system of equations:
\begin{equation}
\label{eq5} \frac{\ddot{B}}{B} + \frac{\ddot{C}}{C} + \frac{\dot{B}\dot{C}}{BC} = - \omega \rho,
\end{equation}
\begin{equation}
\label{eq6} \frac{\ddot{C}}{C} + \frac{\ddot{A}}{A} + \frac{\dot{C}\dot{A}}{CA}= - (\omega + \delta)\rho ,
\end{equation}
\begin{equation}
\label{eq7} \frac{\ddot{A}}{A} + \frac{\ddot{B}}{B} + \frac{\dot{A}\dot{B}}{AB} = - (\omega + \gamma)\rho,
\end{equation}
\begin{equation}
\label{eq8}
\frac{\dot{A}\dot{B}}{AB} + \frac{\dot{B}\dot{C}}{BC} + \frac{\dot{C}\dot{A}}{CA} = \rho.
\end{equation}
Here and in what follows an over dot denotes ordinary differentiation with respect to $t$.\\\\
The spatial volume for the model (\ref{eq1}) is given by
\begin{equation}
\label{eq9} V^{3} = ABC.
\end{equation}
We define $a = (ABC)^{\frac{1}{3}}$ as the average scale factor so that the Hubble's parameter
is anisotropic and may be defined as
\begin{equation}
\label{eq10} H = \frac{\dot{a}}{a} = \frac{1}{3}\left(\frac{\dot{A}}{A} + \frac{\dot{B}}{B} + \frac{\dot{C}}{C}\right).
\end{equation}
We define the generalized mean Hubble's parameter $H$ as
\begin{equation}
\label{eq11} H = \frac{1}{3}(H_{x} + H_{y} + H_{z}),
\end{equation}
where $H_{x}$ = $\frac{\dot{A}}{A}$, $H_{y} = \frac{\dot{B}}{B}$ and $H_{z} = \frac{\dot{C}}{C}$  are
the directional Hubble's parameters in the directions of x, y and z respectively. \\\\
An important observational quantity is the deceleration parameter $q$, which is defined as
\begin{equation}
\label{eq12} q = - \frac{a\ddot{a}}{\dot{a}^{2}}.
\end{equation}
The scalar expansion $\theta$, shear scalar $\sigma^{2}$ and the average anisotropy parameter $A_{m}$
are defined by
\begin{equation}
\label{eq13}
\theta = \frac{\dot{A}}{A} + \frac{\dot{B}}{B} + \frac{\dot{C}}{C},
\end{equation}
\begin{equation}
\label{eq14}
\sigma^{2} = \frac{1}{2}\left(\sum^{3}_{i = 1}H^{2}_{i} - \frac{1}{3}\theta^{2}\right),
\end{equation}
\begin{equation}
\label{eq15} A_{m} = \frac{1}{3}\sum_{i = 1}^{3}{\left(\frac{\triangle
H_{i}}{H}\right)^{2}},
\end{equation}
where $\triangle H_{i} = H_{i} - H (i = 1, 2, 3)$.
\section{Solutions of the Field Equations}
The field equations (\ref{eq5})-(\ref{eq8}) are a system of four equations with seven unknown parameters 
$A$, $B$, $C$, $\rho$, $\omega$, $\delta$ and $\gamma$. Three additional constraints relating these parameters 
are required to obtain explicit solutions of the system. \\\\
Firstly we apply the special law of variation for generalized Hubble's parameter that yields a constant value of 
deceleration parameter. Since the line element (\ref{eq1}) is completely characterized by Hubble parameter $H$, 
therefore, let us consider that the mean Hubble parameter $H$ is related to the average scale factor $a$ by 
the relation
\begin{equation}
\label{eq16} H = \ell a^{-n} = \ell (ABC)^{-\frac{n}{3}},
\end{equation}
where $\ell (> 0)$ and $ n (\geq 0)$ are constants. Such type of relations have already been considered by Berman 
\cite{ref60}, Berman and Gomide \cite{ref61} for solving FRW models. Latter on many authors (see, Singh et al. 
\cite{ref62}$-$\cite{ref64}, Singh and Baghel \cite{ref65}, Pradhan and Jotania \cite{ref66} and references therein) 
have studied flat FRW and Bianchi type models by using the special law for Hubble parameter that yields constant 
value of deceleration parameter. \\\\
From (\ref{eq10}) and (\ref{eq16}), we get
\begin{equation}
\label{eq17} \dot{a} = \ell a^{-n+1},
\end{equation}
and
\begin{equation}
\label{eq18} \ddot{a} = - \ell^{2}(n - 1)a^{-2n + 1}.
\end{equation}
Substituting (\ref{eq17}) and (\ref{eq18}) into (\ref{eq12}), we get
\begin{equation}
\label{eq19} q = n - 1.
\end{equation}
We observe that the relation (\ref{eq19}) gives $q$ as a constant. The sign of $q$ indicated whether the model 
inflates or not. The positive sign of $q$ i.e. $(n > 1)$ correspond to ``standard" decelerating model whereas 
the negative sign of $q$ i.e. $0 \leq n < 1$ indicates inflation. It is remarkable to mention here that though 
the current observations of SNe Ia and CMBR favours accelerating models ($ q < 0$), but both do not altogether 
rule out the decelerating ones which are also consistent with these observations (see, Vishwakarma \cite{ref67}). \\\\
Integrating Eq. (\ref{eq17}) we obtain the law of average scale factor `a' as
\begin{equation}
\label{eq20} a = (n \ell t + c_{1})^{\frac{1}{n}}  ~ ~ \mbox{for} ~ ~   n\neq 0 ,
\end{equation}
and
\begin{equation}
\label{eq21} a = c_{2}e^{\ell t}  ~ ~ \mbox{for} ~ ~  n = 0,
\end{equation}
where $c_{1}$ and $c_{2}$ are constants of integration. Thus, the law (\ref{eq16}) provides two types of the
expansion in the universe i.e., (i) power-law (\ref{eq20}) and (ii) exponential-law (\ref{eq21}). \\\\
Secondly we assume that the component $\sigma^{1}_{~1}$ of the shear tensor $\sigma^{j}_{~i}$ is proportional to the
expansion scalar ($\theta$), i.e., $\sigma^{1}_{~1} \propto \theta$. This condition leads to the following relation
between the metric potentials:
\begin{equation}
\label{eq22} A = (BC)^{m},
\end{equation}
where $m$ is a positive constant. The motive behind assuming this condition is explained with
reference to Thorne \cite{ref68}, the observations of the velocity-red-shift relation for extragalactic 
sources suggest that Hubble expansion of the universe is isotropic today within $\approx 30$ per cent 
(Kantowski and Sachs \cite{ref69}; Kristian and Sachs \cite{ref70}). To put more precisely, red-shift studies 
place the limit
$$
\frac{\sigma}{H} \leq 0.3,
$$
on the ratio of shear $\sigma$ to Hubble constant $H$ in the neighbourhood of our Galaxy today. Collins 
et al. \cite{ref71} have pointed out that for spatially homogeneous metric, the normal congruence to the 
homogeneous expansion satisfies that the condition $\frac{\sigma}{\theta}$ is constant. \\\\ 
Thirdly we assume that the deviations from $\omega$ along y and z axes are same i.e. 
$\gamma = \delta$. \\\\
Subtracting (\ref{eq6}) from (\ref{eq7}), and taking integral of the resulting equation two times, we get
\begin{equation}
\label{eq23}
\frac{B}{C} = c_{3} \exp \left[c_{4} \int (ABC)^{-1} dt\right],
\end{equation}
where $ c_{3} $ and  $c_{4} $ are constants of integration. \\
In the following Subsections (3.1) \& (3.2), we consider two types of solutions, i.e., for power-law and 
exponential form volumetric expansion.
\subsection{Case (i): When $n \ne 0$ i.e. model for power-law expansion}
After solving the field equations (\ref{eq6}) - (\ref{eq8}) for the power-law volumetric expansion (\ref{eq20}) 
by considering Eqs. (\ref{eq22}) and (\ref{eq23}), we obtain the scale factors as follows 
\begin{equation}
\label{eq24}
A(t) = (n \ell t + c_{1})^{\frac{3m}{n(m + 1)}},
\end{equation}
\begin{equation}
\label{eq25}
B(t) = \sqrt{c_{3} }(n \ell t + c_{1})^{\frac{3}{2n(m + 1)}} \exp{\left[\frac{c_{4} }{2\ell (n - 3)}
(n\ell t + c_{1})^{\frac{n - 3}{n} }\right]},
\end{equation}
\begin{equation}
\label{eq26}
C(t) = \frac{1}{\sqrt{c_{3} }}(n\ell t + c_{1})^{\frac{3}{2n(m + 1)}}\exp{\left[-\frac{c_{4} }{2\ell(n - 3)}
(n\ell t + c_{1})^{\frac{n - 3}{n}} \right]},
\end{equation}
provided $n \neq 3$.\\\\
Hence the model (\ref{eq1}) is reduced to
\[
 ds^{2} = - dt^{2} +  (n \ell t + c_{1})^{\frac{6m}{n(m + 1)}}dx^{2} + c_{3} (n \ell t + c_{1})^{\frac{3}
{n(m + 1)}} \exp{\left[\frac{c_{4} }{\ell (n - 3)} (n\ell t + c_{1})^{\frac{n - 3}{n} }\right]} dy^{2}
\]
\begin{equation}
\label{eq27}
+ \frac{1}{c_{3}}(n\ell t + c_{1})^{\frac{3}{n(m + 1)}} \exp{\left[-\frac{c_{4} }{\ell(n-3)}
(n\ell t + c_{1})^{\frac{n-3}{n} }\right]}dz^{2}.
\end{equation}
The rate of expansion $H_{i}$ in the direction of x, y and z read as
\begin{equation}
\label{eq28}
H_{x} = \frac{\dot{A}}{A} = \frac{3m \ell}{m + 1}(n\ell t + c_{1})^{-1},
\end{equation}
\begin{equation}
\label{eq29}
H_{y} = \frac{\dot{B}}{B} = \frac{3\ell}{2(m + 1)}(n\ell t + c_{1})^{-1} + \frac{1}{2}c_{4}(n\ell t + c_{1})^
{-\frac{3}{n}},
\end{equation}
\begin{equation}
\label{eq30}
H_{z} = \frac{\dot{C}}{C} = \frac{3\ell}{2(m + 1)}(n\ell t + c_{1})^{-1} -\frac{1}{2}c_{4}
(n\ell t + c_{1})^{-\frac{3}{n}}.
\end{equation}
Thus the Hubble's parameter $H$, scalar of expansion $\theta$, shear scalar $\sigma$ and the average anisotropy 
parameter $A_{m}$ are given by
\begin{equation}
\label{eq31}
\theta = 3H = 3\ell(n\ell t + c_{1})^{-1},
\end{equation}
\begin{equation}
\label{eq32}
\sigma^{2} = \frac{3\ell^{2}(2m - 1)^{2}}{2(m + 1)^{2}}(n\ell t + c_{1})^{-2} + \frac{1}{4}c_{4}^{2}
(n\ell t + c_{1})^{-\frac{6}{n}},
\end{equation}
\begin{equation}
\label{eq33}
A_{m} = \frac{(2m - 1)^{2}}{(m + 1)^{2}} + \frac{c_{4}^{2}}{6\ell^{2}}(n \ell t + c_{1})^{\frac{2(n - 3)}{n}}.
\end{equation}
\begin{figure}[ht]
\centering
\includegraphics[width=10cm,height=10cm,angle=0]{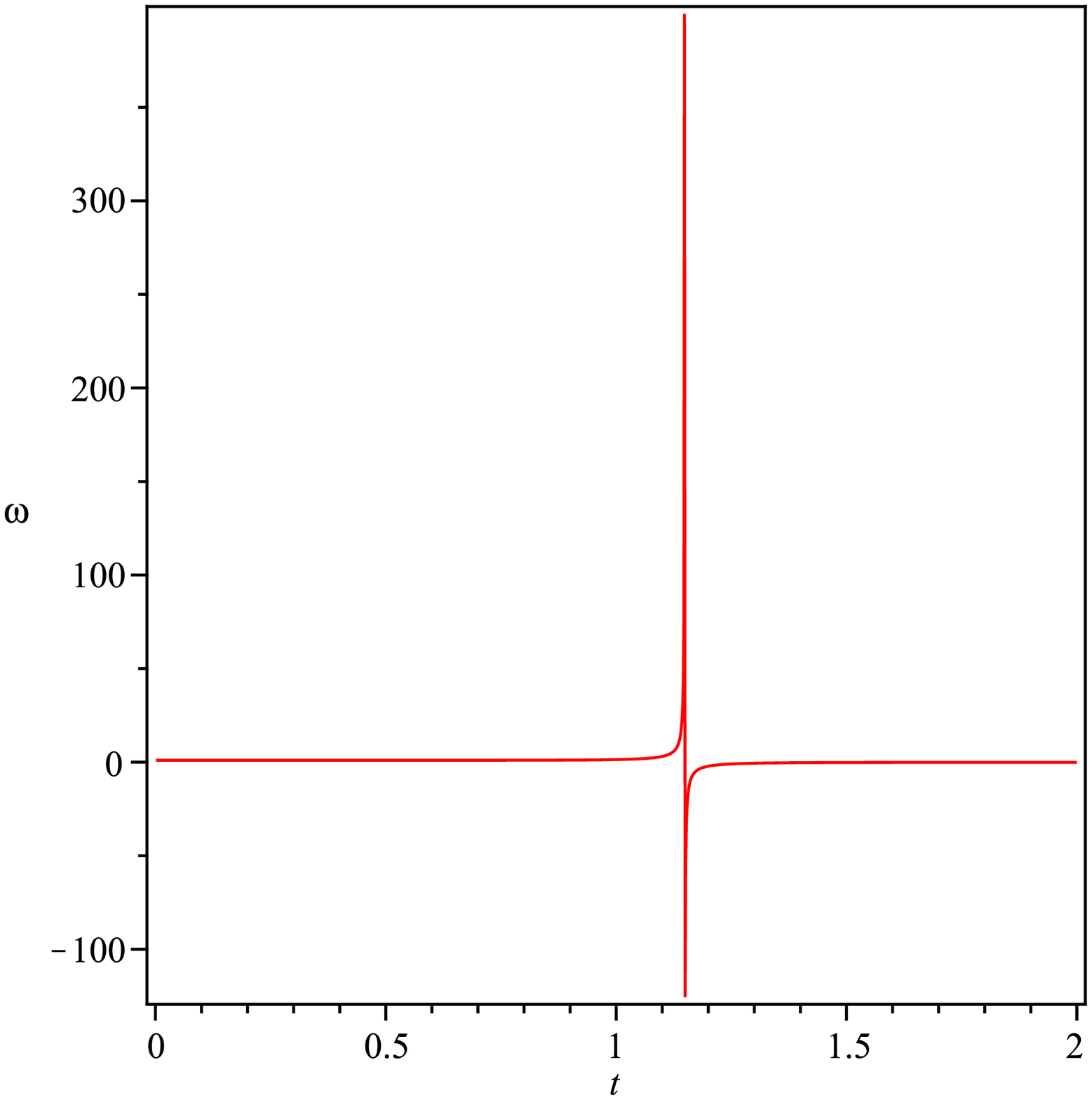} \\
\caption{The plot of EoS parameter $\omega$ versus $t$ in power-law expansion for $q < 0$. Here $\ell = 0.33$, 
$n = 0.5$, $m = 2$, $c_{4} = 2$}.
\end{figure}
\begin{figure}[ht]
\centering
\includegraphics[width=10cm,height=10cm,angle=0]{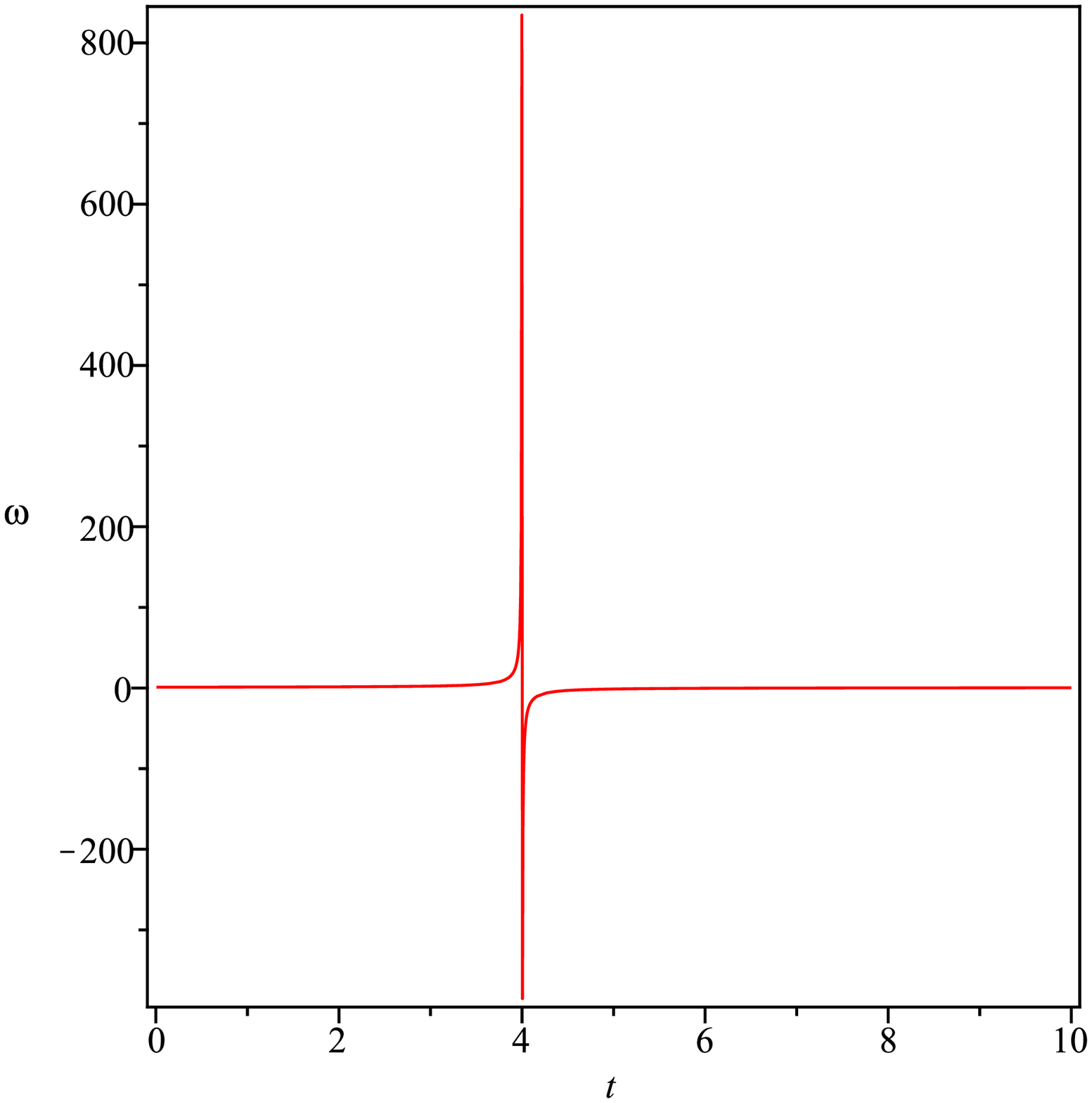} \\
\caption{The plot of EoS parameter $\omega$ versus $t$ in power-law expansion for $q > 0$. Here $\ell = 0.33$, 
$n = 2$, $m = 2$, $c_{4} = 2$}.
\end{figure}
\begin{figure}[ht]
\centering
\includegraphics[width=10cm,height=10cm,angle=0]{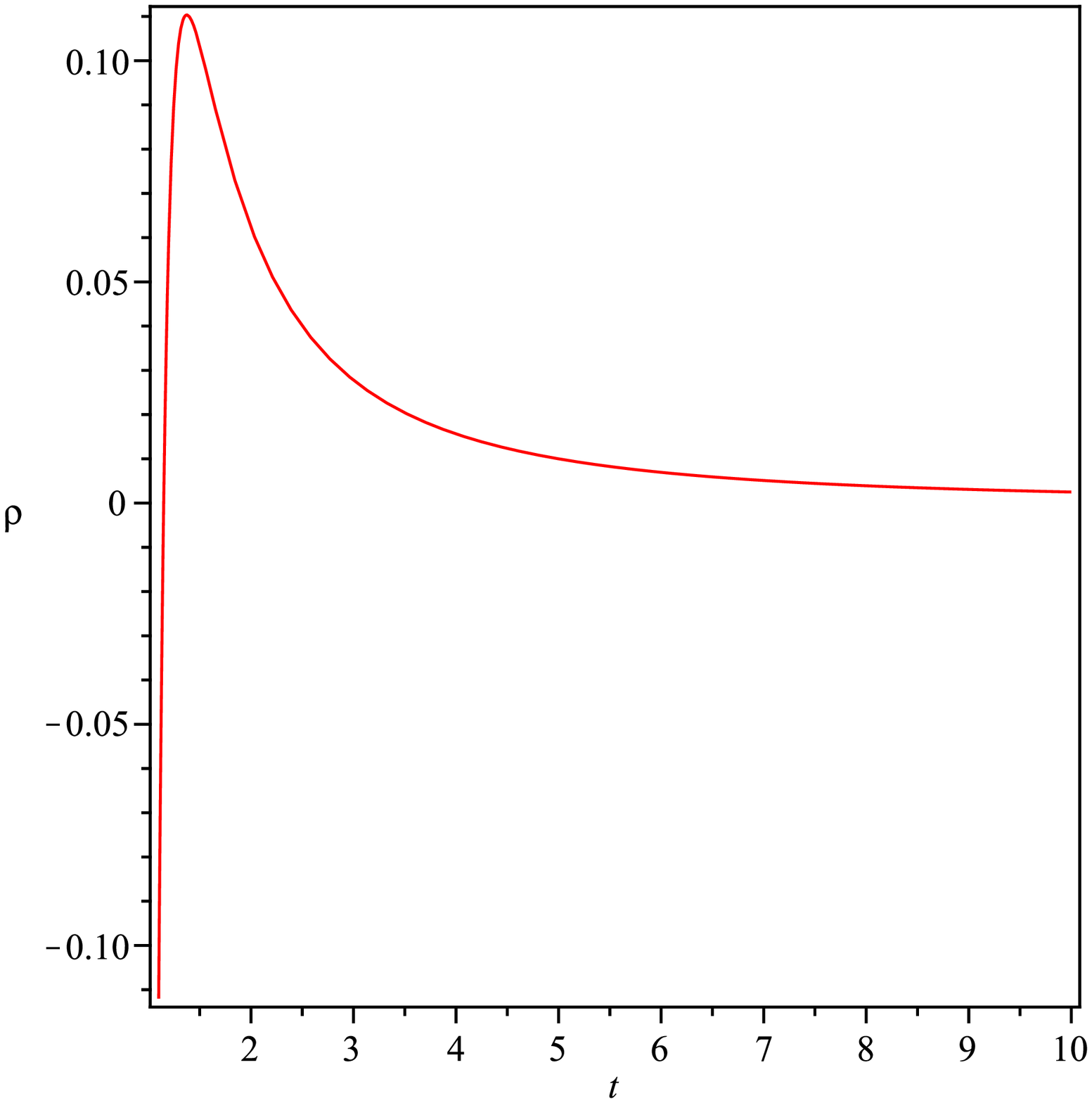} \\
\caption{The plot of energy density $\rho$ versus $t$ in power-law expansion for $q < 0$. 
Here $\ell = 0.33$, $n = 0.5$, $m = 2$, $c_{4} = 2$}.
\end{figure}
\begin{figure}[ht]
\centering
\includegraphics[width=10cm,height=10cm,angle=0]{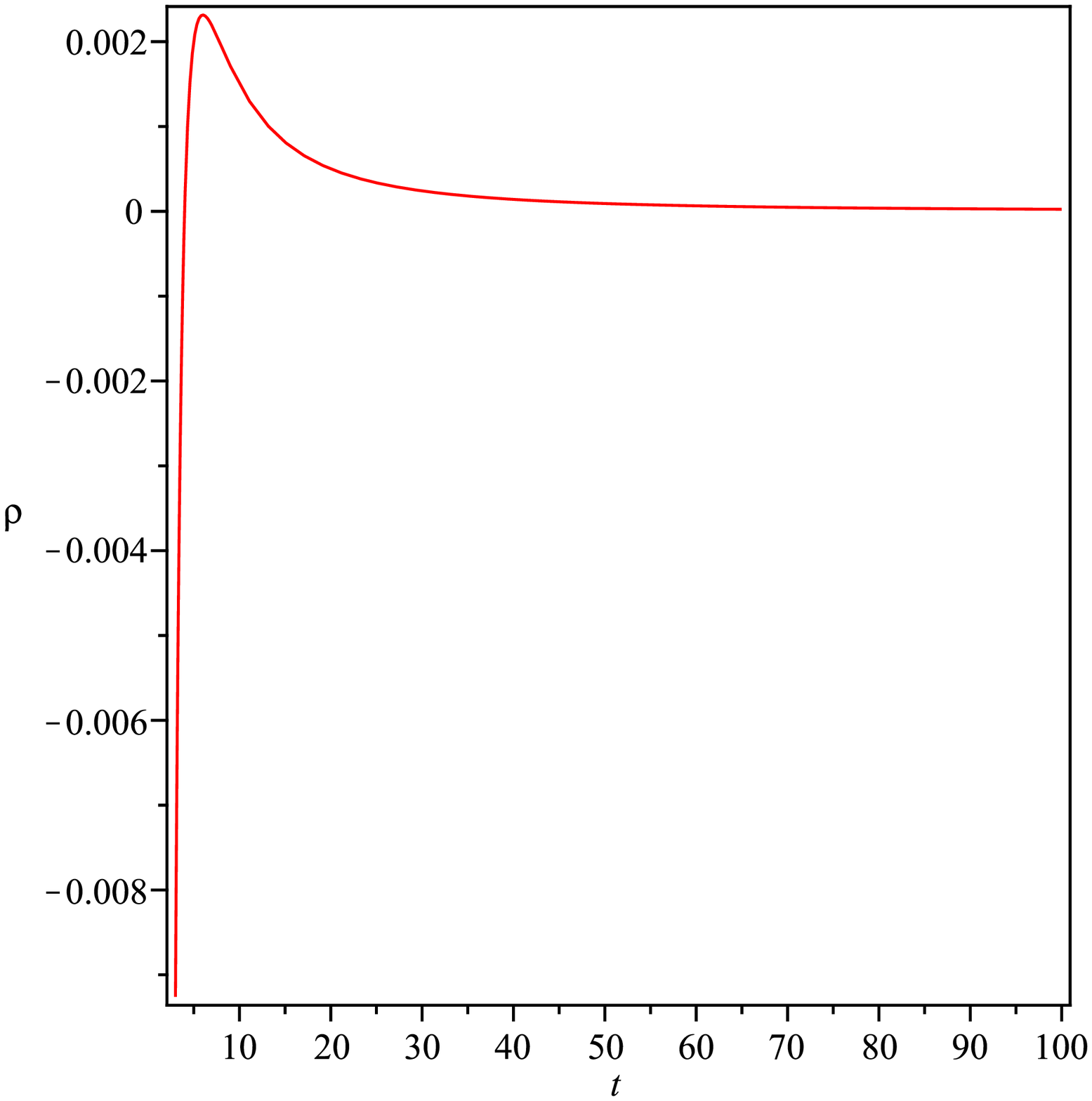} \\
\caption{The plot of energy density $\rho$ versus $t$ in power-law expansion for $q > 0$. 
Here $\ell = 0.33$, $n = 2$, $m = 2$, $c_{4} = 2$}.
\end{figure}
\begin{figure}[ht]
\centering
\includegraphics[width=10cm,height=10cm,angle=0]{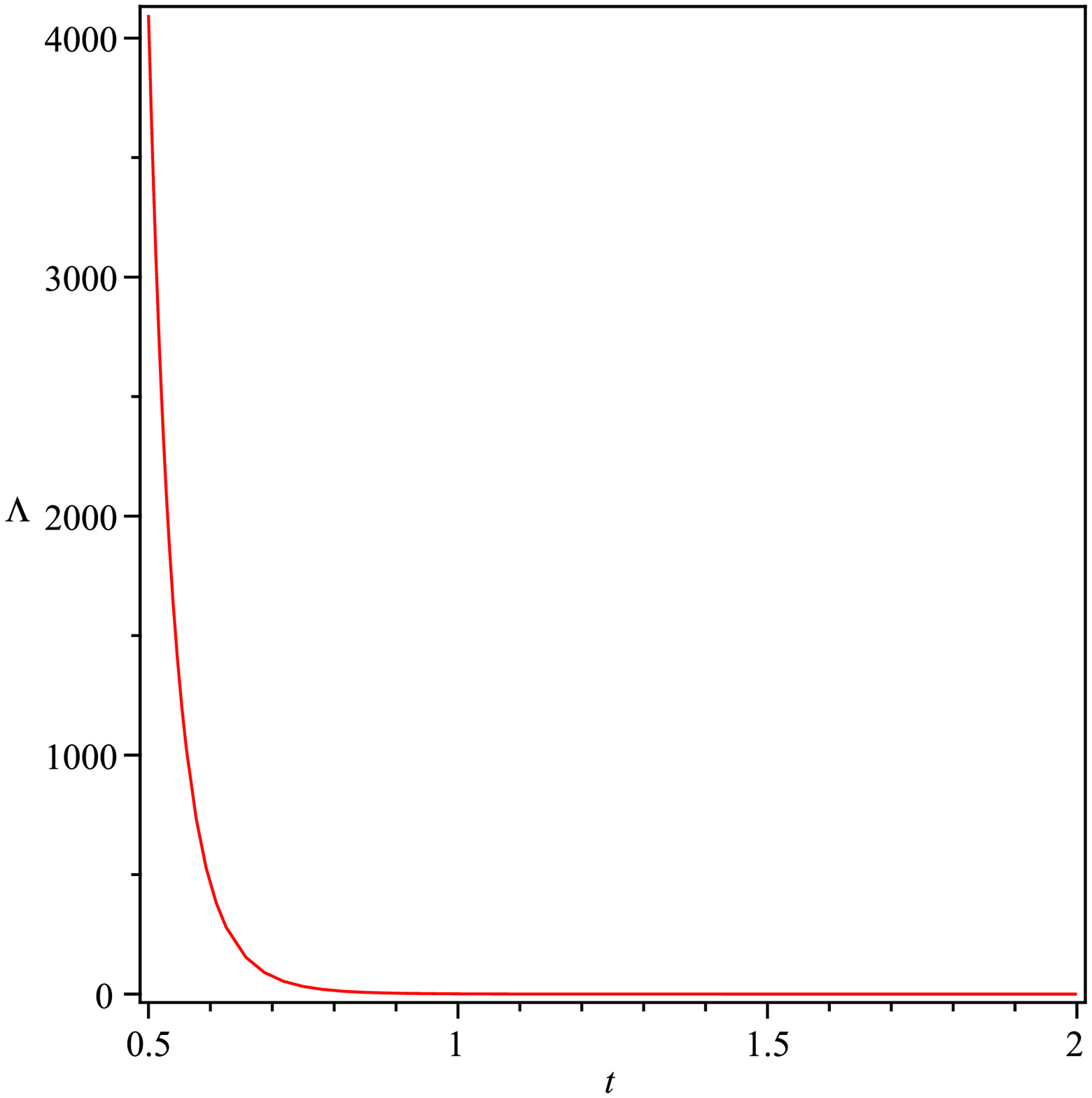} \\
\caption{The plot of cosmological constant $\Lambda$ versus $t$ in power-law expansion for $q < 0$. 
Here $\ell = 0.33$, $n = 0.5$, $m = 2$, $c_{4} = 2$}.
\end{figure}
\begin{figure}[ht]
\centering
\includegraphics[width=10cm,height=10cm,angle=0]{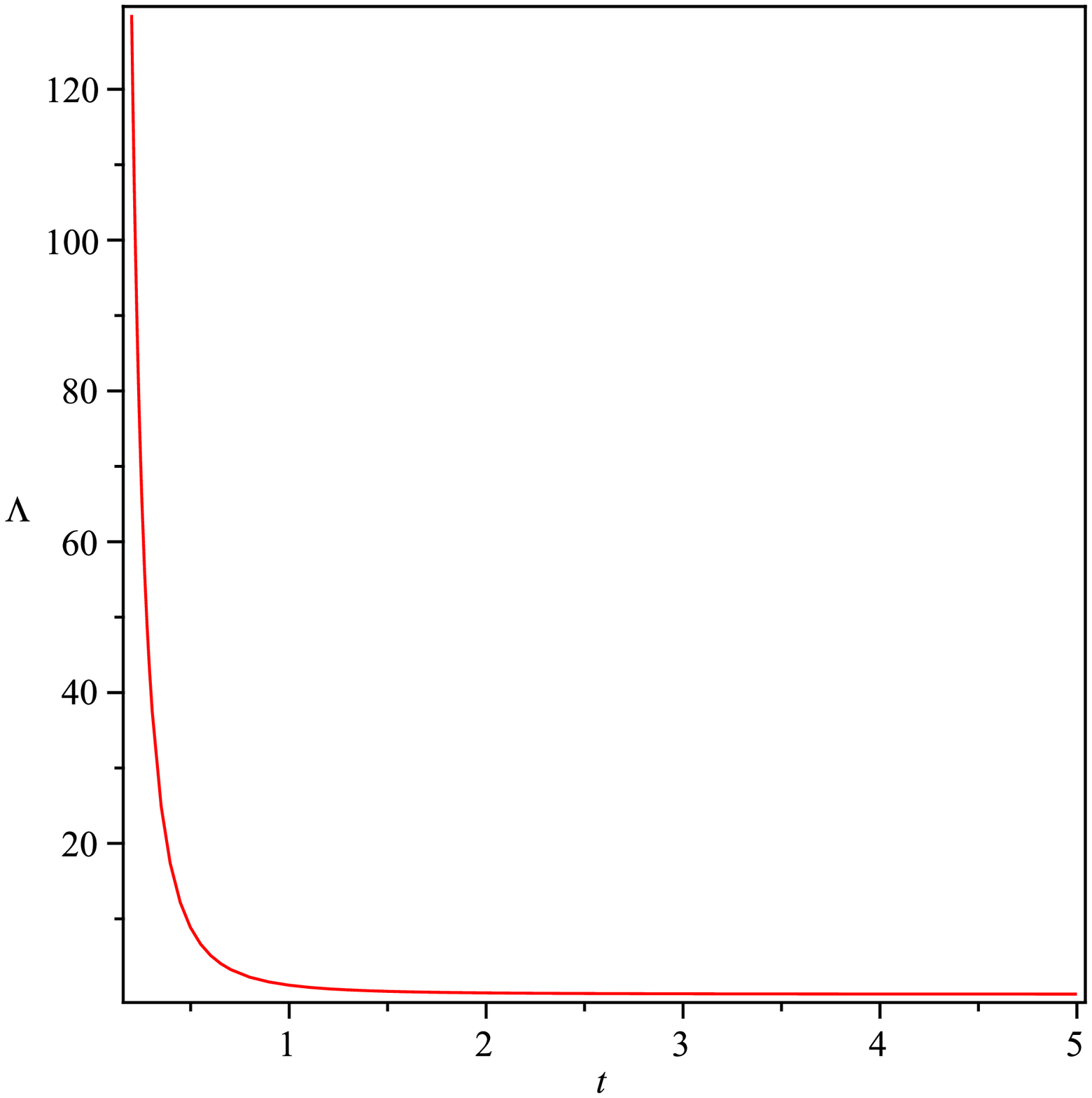} \\
\caption{The plot of cosmological constant $\Lambda$ versus $t$ in power-law expansion for $q > 0$. 
Here $\ell = 0.33$, $n = 2$, $m = 2$, $c_{4} = 2$}.
\end{figure}

The energy density of the fluid can be find by using Eqs. (\ref{eq24}) - (\ref{eq26}) in (\ref{eq8})
\begin{equation}
\label{eq34}
\rho = \frac{9\ell^{2}(4m + 1)}{4(m + 1)^{2}}(n\ell t + c_{1})^{-2}-\frac{1}{4}c_{4}^{2}(n\ell t + c_{1})^
{-\frac{6}{n}}.
\end{equation}
Using Eqs. (\ref{eq25}), (\ref{eq26}) and (\ref{eq34}) in (\ref{eq5}), the EoS parameter $\omega$ is obtained as
\begin{equation}
\label{eq35}
\omega = \frac{3\ell^{2}\left[\frac{-4n(m + 1) + 9}{4(m + 1)^{2}}\right](n\ell t + c_{1})^{-2} + 
\frac{1}{4}c^{2}_{4}(n\ell t + c_{1})^{-\frac{6}{n}}}{\frac{1}{4}c^{2}_{4}(n\ell t + c_{1})^{-\frac{6}{n}} - 
\frac{9\ell^{2}(4m + 1)}{4(m + 1)^{2}}(n\ell t + c_{1})^{-2}}.
\end{equation}
Using Eqs. (\ref{eq24})-(\ref{eq26}), (\ref{eq34}) and (\ref{eq35}) in either (\ref{eq6}) or (\ref{eq7}), the 
skewness parameters $\delta$  and $\gamma$ (i.e. deviation from $\omega$ along y-axis and z-axis) are computed 
as
\begin{equation}
\label{eq36}
\delta = \gamma = \frac{\frac{3\ell^{2}(3 - n)[4m n(m + 1) - 3]}{4n(m + 1)^{2}}(n\ell t + c_{1})^{-2}}{\frac{1}{4}
c_{4}^{2}(n\ell t + c_{1})^{-\frac{6}{n}} - \frac{9\ell^{2}(4m + 1)}{4(m + 1)^{2}}(n\ell t + c_{1})^{-2}}.
\end{equation}
From equation (\ref{eq35}), it is observed that the equation of state parameter $\omega$ is time
dependent, it can be function of redshift $z$ or scale factor $V$ as well. The redshift
dependence of $\omega$ can be linear like
\begin{equation}
\label{eq37}
\omega(z) = \omega_{0} + \omega^{1} z,
\end{equation}
with $\omega^{1}$ = $(\frac{d\omega}{dz})_{z} = 0$ (see Refs. Huterer and Turner \cite{ref29}; Weller and Albrecht
\cite{ref30}) or nonlinear as
\begin{equation}
\label{eq38}
\omega(z) = \omega_{0} + \frac{\omega_{1}z}{1 + z},
\end{equation}
(Polarski and Chavellier \cite{ref72}; Linder \cite{ref73}). The SNe Ia data suggests that $-1.67 < \omega < -0.62$
(Knop et al. \cite{ref35}) while the limit imposed on $\omega$ by a combination of SNe Ia data (with CMB anisotropy)
and galaxy clustering statistics is $-1.33 < \omega < -0.79$ (Tegmark et al. \cite{ref36}). So, if the present work
is compared with experimental results mentioned above, then one can conclude that the limit of $\omega$
provided by equation (\ref{eq35}) may accommodated with the acceptable range of EoS parameter. Also it is observed
that at $t=t_{c}$, $\omega$ vanishes, where
$t_{c}$ is a critical time given by
\begin{equation}
\label{eq39} t_{c} = \frac{1}{n\ell}\left(\frac{\ell\sqrt{3[4n( m+ 1)-9]}}{c_{4}(m + 1)}\right)^{\frac{n}{n-3}} - 
\frac{c_{1}}{n\ell}.
\end{equation}
Thus for this particular time our model represents a dusty universe.\\\\
For the value of $\omega$ to be in consistent with observation (Knop et al. \cite{ref35}), we have the following
general condition
\begin{equation}
\label{eq40}
t_{1} < t < t_{2},
\end{equation}
where
\begin{equation}
\label{eq41}
t_{1} = \frac{1}{n\ell}\left[1.12\ell\frac{\sqrt{4n(m + 1) + 5(4m + 1)}}{c_{4}(m + 1)}\right]^{\frac{n}{n - 3}} - 
\frac{c_{1}}{n\ell},
\end{equation}
and
\begin{equation}
\label{eq42}
t_{2} = \frac{1}{n\ell}\left[1.36\ell\frac{\sqrt{4n(m + 1) + 7.44m - 7.14}}{c_{4}(m + 1)}\right]^{\frac{n}{n-3}} - 
\frac{c_{1}}{n\ell}.
\end{equation}
For this constrain we obtain  $-1.67 < \omega < -0.62$ which is in good agreement with the limit obtained from
observational results coming from SNe Ia data (Knop et al. \cite{ref35} ). We note that the earlier real matter at 
$t \leq t_{c}$, where $\omega \geq 0$ later on at $t > t_{c}$, where $\omega < 0$ converted
to the dark energy dominated phase of universe.\\\\
We also observe that if
\begin{equation}
\label{eq43}
t_{0} = \frac{1}{n\ell}\left[1.22\ell\frac{\sqrt{4n(m + 1) + 6(2m - 1)}}{c_{4}(m + 1)}\right]^{\frac{n}{n - 3}} - 
\frac{c_{1}}{n\ell}.
\end{equation}
for $t = t_{0}$, $\omega = -1$ (i.e. cosmological constant dominated universe), and when $t < t_{0}$, $\omega > -1$ 
(i.e. quintessence), and for $t > t_{0}$, $\omega < -1$ (i.e. super quintessence or phantom fluid dominated universe) 
(Caldwell  \cite{ref74}).\\\\
{\bf Figures 1 \& 2} depict the variation of equation of state parameter ($\omega$) versus cosmic time ($t$) in 
the two modes (i.e. $q < 0$ and $q > 0$) of evolution of the universe, as a representative case with appropriate 
choice of constants of integration and other physical parameters using reasonably well known situations. From 
{\bf Figure 1 \& 2}, we conclude that in early stage of evolution of the universe, the EoS parameter $\omega$ was 
positive (i.e. the universe was matter dominated) and at late time it is evolving with negative value (i.e. at 
the present time). The earlier real matter later on converted to the dark energy dominated phase of the universe 
in both accelerating \& decelerating modes.\\\\
From Eq. (\ref{eq34}), we note that energy density of the fluid $\rho(t)$ is a decreasing function of time and 
$\rho \geq 0$ when
\begin{equation}
\label{eq44}
t \geq \frac{1}{n\ell}\left[\left(\frac{c_{4}(m + 1)}{3\ell \sqrt{4m + 1}}\right)^{\frac{n}{3 - n}}
- c_{1} \right].
\end{equation}
{\bf Figures 3 \& 4} are the plots of energy density of the fluid ($\rho$) versus time in accelerating and decelerating 
modes of the universe respectively. In both modes we observe that in early universe $\rho$ is negative and suddenly 
reaches to maximum positive value and then decreases with time and at late time it approaches to zero.\\\\
In absence of any curvature, matter energy density $\Omega_{m}$ and dark energy $\Omega_{\Lambda}$
are related by the equation
\begin{equation}
\label{eq45}
\Omega_{m} + \Omega_{\Lambda} = 1,
\end{equation}
where $\Omega_{m} = \frac{\rho}{3H^{2}}$ and $\Omega_{\Lambda} = \frac{\Lambda}{3H^{2}}$. Thus equation 
(\ref{eq45}), reduces to
\begin{equation}
\label{eq46}
\frac{\rho}{3H^{2}} + \frac{\Lambda}{3H^{2}} = 1.
\end{equation}
Using equations (\ref{eq31}) and (\ref{eq34}), in equation (\ref{eq46}), the cosmological constant is computed and 
obtained as
\begin{equation}
\label{eq47}
\Lambda = \frac{3\ell^{2}(4m^{2} - 4m + 1)}{4(m + 1)^{2}}(n\ell t + c_{1})^{-2} + \frac{1}{4}c_{4}^{2}
(n\ell t + c_{1})^{-\frac{6}{n}}.
\end{equation}
From eq. (\ref{eq47}) we observe that $\Lambda$ is a decreasing function of time and it is always positive for 
$m > \frac{1}{2}$. \\\\
In recent time the $\Lambda$-term has interested theoreticians and observers for various reasons. The 
nontrivial role of the vacuum in the early universe generate a $\Lambda$-term that leads to inflationary 
phase. Observationally, this term provides an additional parameter to accommodate conflicting data on the 
values of the Hubble constant, the deceleration parameter, the density parameter and the age of the universe 
(for example, see the references Gunn and Tinsley \cite{ref75}; Wampler and Burke \cite{ref76}).
The behaviour of the universe in this model will be determined by the cosmological term $\Lambda$, this term
has the same effect as a uniform mass density $\rho_{eff} = - \Lambda $ which is constant in time. A positive 
value of $\Lambda$ corresponds to a negative effective mass density (repulsion). Hence, we expect that in the 
universe with a positive value of $\Lambda$ the expansion will tend to accelerate whereas in the universe with 
negative value of $\Lambda$ the expansion will slow down, stop and reverse. In a universe with both matter and 
vacuum energy, there is a competition between the tendency of $\Lambda$ to cause acceleration and the tendency 
of matter to cause deceleration with the ultimate fate of the universe depending on the precise amounts of each 
component. This continues to be true in the presence of spatial curvature, and with a nonzero cosmological constant 
it is no longer true that the negatively curved (``open'') universes expand indefinitely while positively curved 
(``closed'') universes will necessarily re-collapse - each of the four combinations of negative or positive curvature 
and eternal expansion or eventual re-collapse become possible for appropriate values of the parameters. There may 
even be a delicate balance, in which the competition between matter and vacuum energy is needed drawn and the 
universe is static (non expanding). { \it The search for such a solution was Einstein's original motivation for 
introducing the cosmological constant.} \\\\
{\bf Figures 5 \& 6} are the plots of cosmological constant $\Lambda$ versus time in accelerating and decelerating 
mode of the universe respectively. In both mode we observe that cosmological parameter is decreasing function of time 
and it approaches a small positive value at late time (i.e. at present epoch). Recent cosmological observations 
(Garnavich et al. \cite{ref1,ref2}; Perlmutter et al. \cite{ref3}$-$ \cite{ref5}; Riess et al. \cite{ref6,ref77,ref78}; 
Schmidt et al. \cite{ref7}) suggest the existence of a positive cosmological constant $\Lambda$ with the 
magnitude $\Lambda(G\hbar/c^{3})\approx 10^{-123}$. These observations on magnitude and red-shift of type Ia 
supernova suggest that our universe may be an accelerating one with induced cosmological density through the 
cosmological $\Lambda$-term. But this does not rule out the decelerating ones which are also consistent with 
these observations (Vishwakarma \cite{ref67}). Thus the nature of $\Lambda$ in our derived DE model is supported 
by recent observations. \\\\
From Eqs. (\ref{eq28}) - (\ref{eq33}), it can be seen that the spatial volume is zero at $t = -\frac{c_{1}}{n\ell}$,
and it increases with the cosmic time. The parameters $H_{i}$, H, $\theta$ and $\sigma$ diverge at the
initial singularity. There is a Point Type singularity (MacCallum \cite{ref79}) at $ t = -\frac{c_{1}}{n \ell}$ in 
the model. The mean anisotropic parameter is an increasing function of time for $n > 3$ whereas
for $n < 3$ it decreases with time. Thus, the dynamics of the mean anisotropy parameter depends on the
value of n. Since $\frac{\sigma^{2}}{\theta^{2}} = $ constant (from early to late time), the model does
not approach isotropy through the whole evolution of the universe.
\subsection{Case (ii): When $n = 0$ i.e model for exponential expansion}
After solving the field equations (\ref{eq6}) - (\ref{eq8}) for the power-law volumetric expansion (\ref{eq21}) 
by considering Eqs. (\ref{eq22}) and (\ref{eq23}), we obtain the scale factors as follows
\begin{equation}
\label{eq48}
A(t) = c_{2}^{\frac{3m}{m + 1}}\exp{\left(\frac{3m\ell}{m + 1}t\right)},
\end{equation}
\begin{equation}
\label{eq49}
B(t) = \sqrt{c_{3}}c_{2}^{\frac{3}{2(m + 1)}}\exp{\left[\frac{3\ell}{2(m + 1)}t - \frac{c_{4}}{6\ell c_{2}^{3}}
e^{-3\ell t}\right]},
\end{equation}
\begin{equation}
\label{eq50}
C(t) = \frac{c_{2}^{\frac{3}{2(m + 1)}}}{\sqrt{c_{3}}}\exp{\left[\frac{3\ell}{2(m + 1)}t +
\frac{c_{4}}{6\ell c_{2}^{3}}e^{-3\ell t}\right]}.
\end{equation}
Hence the model (\ref{eq1}) is reduced to
\[
 ds^{2} = - dt^{2} + c_{2}^{\frac{6m}{m + 1}}\exp{\left(\frac{6m\ell}{m+1}t\right)} dx^{2} +
c_{3}c_{2}^{\frac{3}{(m + 1)}}\exp{\left[\frac{3\ell}{(m + 1)}t - \frac{c_{4}}{3\ell c_{2}^{3}}
e^{-3\ell t}\right]}dy^{2}
\]
\begin{equation}
\label{eq51}
+ \frac{c_{2}^{\frac{3}{(m + 1)}}}{c_{3}}\exp{\left[\frac{3\ell}{(m + 1)}t +
\frac{c_{4}}{3\ell c_{2}^{3}}e^{-3\ell t}\right]}dz^{2}.
\end{equation}
\begin{figure}[ht]
\centering
\includegraphics[width=10cm,height=10cm,angle=0]{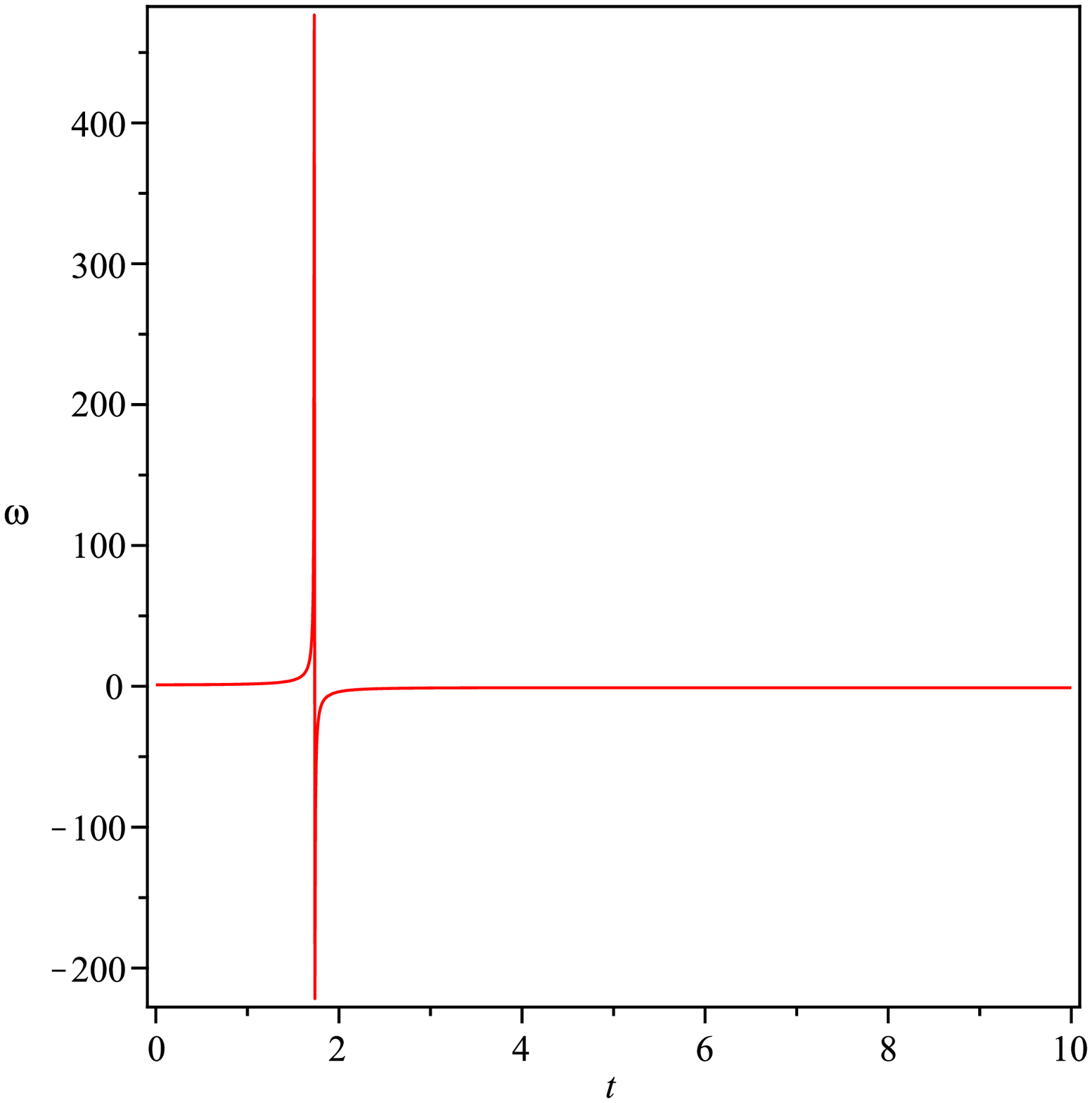} \\
\caption{The plot of EoS parameter $\omega$ versus $t$ in exponentialr-law expansion for $q = -1 0$. 
Here $\ell = 0.33$, $n = 0$, $m = 2$, $c_{4} = 2$}.
\end{figure}
\begin{figure}[ht]
\centering
\includegraphics[width=10cm,height=10cm,angle=0]{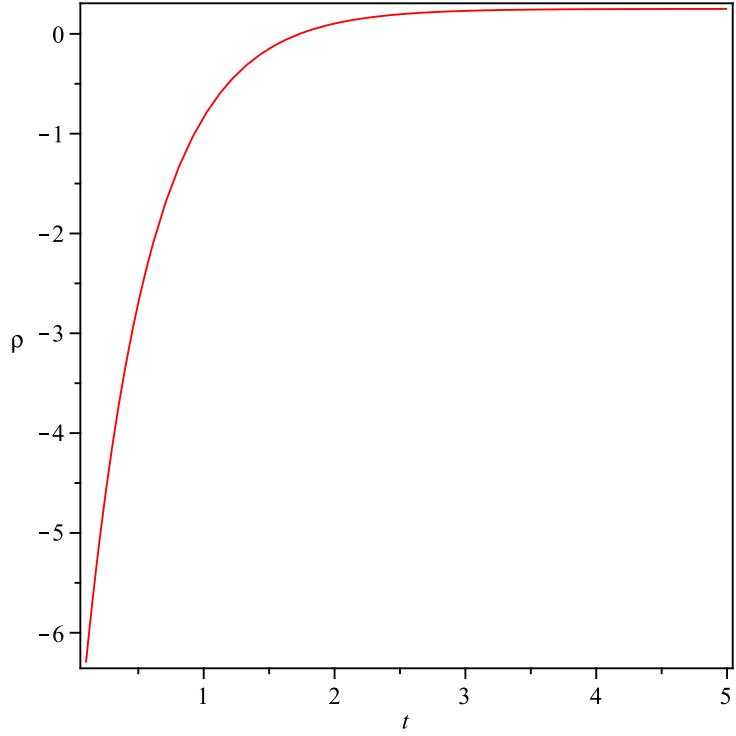} \\
\caption{The plot of energy density $\rho$ versus $t$ in exponential-law expansion for $q = -1$. 
Here $\ell = 0.33$, $n = 0$, $m = 2$, $c_{4} = 2$}.
\end{figure}
\begin{figure}[ht]
\centering
\includegraphics[width=10cm,height=10cm,angle=0]{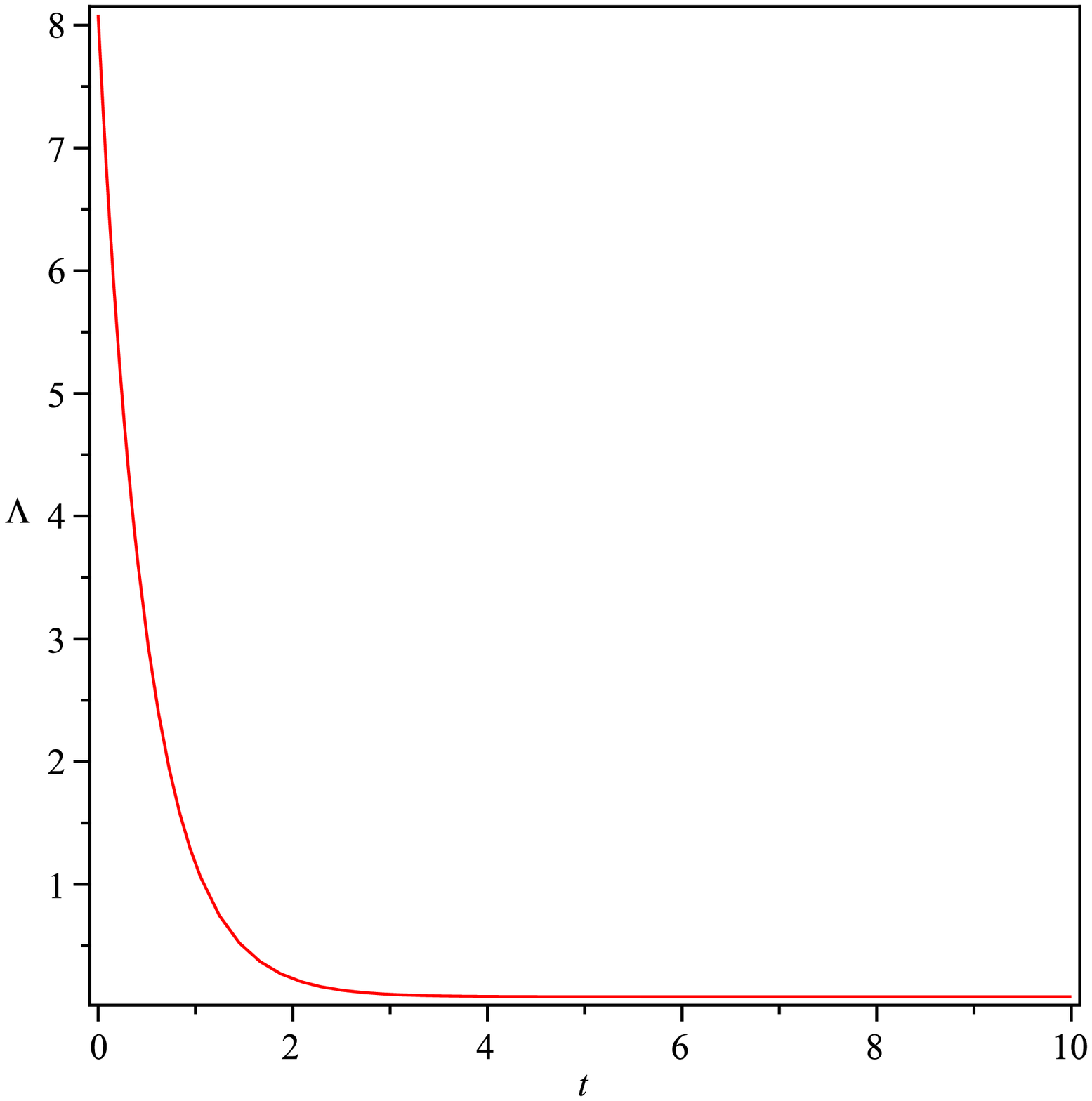} \\
\caption{The plot of cosmological constant $\Lambda$ versus $t$ in exponential-law expansion for $q = -1$. 
Here $\ell = 0.33$, $n = 0$, $m = 2$, $c_{4} = 2$}.
\end{figure}
The rate of expansion $H_{i}$ in the direction of x, y and z and the Hubble parameter are obtained as
\begin{equation}
\label{eq52}
H_{x} = \frac{3m\ell}{m+1},
\end{equation}
\begin{equation}
\label{eq53}
H_{y} = \frac{3\ell}{2(m+1)} + \frac{c_{4}}{2c_{2}^{3}}e^{-3\ell t},
\end{equation}
\begin{equation}
\label{eq54}
H_{z} = \frac{3\ell}{2(m+1)} -\frac{c_{4}}{2c_{2}^{3}}e^{-3\ell t}.
\end{equation}
Thus the Hubble's parameter $H$, scalar of expansion $\theta$, shear scalar $\sigma$ and the average anisotropy 
parameter $A_{m}$ are given by
\begin{equation}
\label{eq55}
H = \ell,
\end{equation}
\begin{equation}
\label{eq56}
\theta = 3\ell,
\end{equation}
\begin{equation}
\label{eq57}
\sigma^{2} = \frac{3\ell^{2}(2m-1)^{2}}{4(m + 1)^{2}} - \frac{c_{4}^{2}}{4c_{2}^{6}}e^{-6\ell t},
\end{equation}
\begin{equation}
\label{eq58}
A_{m} = \frac{(2m-1)^{2}}{2(m + 1)^{2}} - \frac{c_{4}^{2}}{6\ell^{2}c_{2}^{6}}e^{-6\ell t}.
\end{equation}
Using Eqs. (\ref{eq48})-(\ref{eq50}) in (\ref{eq8}), the energy density of the fluid is obtained as
\begin{equation}
\label{eq59}
\rho = \frac{9\ell^{2}(4m + 1)}{4(m + 1)^{2}} - \frac{c_{4}^{2}}{4c_{2}^{6}}e^{-6\ell t}.
\end{equation}
Using Eqs. (\ref{eq49}), (\ref{eq50}) and (\ref{eq59}) in (\ref{eq5}), the EoS parameter $\omega$ is obtained as
\begin{equation}
\label{eq60}
\omega = \frac{\frac{27\ell^{2}}{4(m + 1)^{2}} + \frac{c^{2}_{4}}{4c^{6}_{2}}e^{-6\ell t}}
{\frac{c_{4}^{2}}{4c_{2}^{6}}e^{-6\ell t} - \frac{9\ell^{2}(4m + 1)}{4(m + 1)^{2}}}.
\end{equation}
Using (\ref{eq48})-(\ref{eq50}) and (\ref{eq59}), (\ref{eq60}) in either (\ref{eq6}) or (\ref{eq7}), the skewness 
parameters $\delta$  and $\gamma$ are computed as
\begin{equation}
\label{eq61}
\delta = \gamma = \frac{\frac{9\ell^{2}(2m^{2} + m - 1)}{2(m + 1)^{2}}}{\frac{9\ell^{2}(4m + 1)}{4(m + 1)^{2}} - 
\frac{c_{4}^{2}}{4c_{2}^{6}}e^{-6\ell t}}.
\end{equation}
It is observed that for $t = t_{c}$, $\omega$ vanishes, where $t_{c}$ is a critical time given by
\begin{equation}
\label{eq62}
t_{c} = \frac{1}{6\ell}\ln\left(-\frac{c^{2}_{4}(m + 1)^{2}}{27 c^{6}_{2}\ell^{2}}\right).
\end{equation}
Thus for this particular time our model represent a dusty universe.\\\\
For the value of $\omega$ to be in consistent with observation (Knop et al. \cite{ref35}), we have the following
general condition
\begin{equation}
\label{eq63}
t_{1} < t < t_{2},
\end{equation}
where
\begin{equation}
\label{eq64}
t_{1} = \frac{1}{6\ell}\ln\left(\frac{c^{2}_{4}(m + 1)^{2}}{3.37c^{6}_{2}\ell^{2}(6.7m - 1.33)}\right),
\end{equation}
and
\begin{equation}
\label{eq65}
t_{2} = \frac{1}{6\ell}\ln\left(\frac{c^{2}_{4}(m + 1)^{2}}{5.6c^{6}_{2}\ell^{2}(2.4m - 2.38)}\right).
\end{equation}
For this constrain we obtain  $-1.67 < \omega < -0.62$, which is in good agreement with the limit obtained from
observational results coming from SNe Ia data (Knop et al. \cite{ref35}). We note that the
earlier real matter at $t \leq t_{c}$, where $\omega \geq 0$, later on at $t> t_{c}$, where $\omega < 0$, converted
to the dark energy dominated phase of universe.\\\\
It is also observed that if
\begin{equation}
\label{eq66}
t_{0}= \frac{1}{6\ell}\ln\left(\frac{c^{2}_{4}(m + 1)^{2}}{9c^{6}_{2}\ell^{2}(m - 1)}\right),
\end{equation}
for $t = t_{0}$, $\omega = -1$ (i.e. cosmological constant dominated universe), and when $t < t_{0}$, $\omega > -1$ 
(i.e. quintessence), and for $t > t_{0}$, $\omega < -1$ (i.e. super quintessence or phantom fluid dominated universe) 
(Caldwell \cite{ref74}).\\\\
The variation of equation of state parameter ($\omega$) with cosmic time ($t$) is clearly shown in {\bf Figures 7},
as a representative case with appropriate choice of constants of integration and other physical parameters using 
reasonably well known situations. From {\bf Figure 7}, we conclude that in early stage of evolution of the universe, 
the EoS parameter $\omega$ was positive (i.e. the universe was matter dominated) and at late time it is evolving 
with negative value (i.e. at the present time). The earlier real matter later on converted to the dark energy 
dominated phase of the universe.\\\\
From (\ref{eq59}), it is observed that $\rho(t)$ is a slowly increasing function of time for large t. When 
$t \to \infty$, $\rho$ becomes a constant. The energy condition $\rho \geq 0$ is satisfied under the condition
\begin{equation}
\label{eq67}
e^{3\ell t} \geq \frac{c_{4}(m + 1)}{3\ell c_{2}^{3}\sqrt{(4m + 1)}}. 
\end{equation}
This behavior of $\rho(t)$ is clearly depicted in {\bf Figure 8}. This means there is no density evolution in this 
set up. A possible reason for no evolution of energy density is that the universe could be much rapid in which 
the matter do not get time to readjust on expanding range or other unknown dominated effect which not incorporated 
in potential functions of this space-time for exponential-law.   \\\\
Using equations (\ref{eq55}) and (\ref{eq59}), in equation (\ref{eq46}), the cosmological constant is obtained as
\begin{equation}
\label{eq68}
\Lambda = \frac{3\ell^{2}(4m^{2} - 4m + 1)}{4(m + 1)^{2}} + \frac{c^{2}_{4}}{4c^{6}_{2}}e^{-6\ell t}.
\end{equation}
From Eq. (\ref{eq68}) we observe that $\Lambda$ is a decreasing function of time and it is always positive for 
$m > \frac{1}{2}$. This behavior of cosmological constant $\Lambda$ is shown in {\bf Figure 9}. Thus, in this case also 
we observe that $\Lambda$ is consistent with recent cosmological observations (Garnavich et al. \cite{ref1,ref2}; 
Perlmutter et al. \cite{ref3}$-$ \cite{ref5}; Riess et al. \cite{ref6,ref77,ref78}; Schmidt et al. \cite{ref7}) \\\\
From above results, it is observed that the physical and kinematic quantities are all constant at $t = 0$.
From Eqs. (\ref{eq52}) - (\ref{eq58}), we observe that the kinematic parameters tend to zero as $t \to \infty$. 
The expansion in the model is uniform throughout the time of evolution. Since $\frac{\sigma^{2}}{\theta^{2}} = $ 
constant (from early to late time), the model does not approach isotropy at any time. The derived model is non-singular.
\section{Concluding Remarks}
Two new anisotropic B-I DE models with variable EoS parameter $\omega$ have been investigated which are 
different from the other author's solutions. In the derived models, $\omega$ is obtained as time varying which 
is consistent with recent observations (Knop et al. \cite{ref35}; Tegmark et al. \cite{ref36}). The proposed 
law of variation for the Hubble's parameter yields a constant value of deceleration parameter. The law of variation 
for Hubble's parameter defined by (\ref{eq16}) for B-I space-time gives two types of cosmologies, 
(i) first form for ($n \ne 0$) gives the solution for positive value of deceleration parameter for  $n > 1$ 
and also negative value of deceleration parameter for $0 \leq n < 1$ indicating the power law expansion of the universe 
whereas (ii) second one (for $n = 0$) gives the solution for negative value of deceleration parameter, which shows 
the exponential expansion of the universe. The power law solution represents the singular model where the spatial 
scale factors and volume scalar vanish at $ t = -\frac{c_{1}}{n \ell}$. All the physical parameters are infinite at 
this initial epoch and tend to zero as $t \to \infty$. There is a Point Type singularity (MacCallum \cite{ref79}) at 
$ t = -\frac{c_{1}}{n \ell}$ in the first model. The exponential solutions represents singularity free model of the 
universe. In both cases, since $\frac{\sigma^{2}}{\theta^{2}} = $ constant (from early to late time), the models do
not approach isotropy through the whole evolution of the universe. \\\\
The main features of the models are as follows:
\begin{itemize}
\item The DE models are based on exact solutions of the Einstein's field equations for the anisotropic B-I 
space-time filled with perfect fluid with variable EoS parameter $\omega$. The exact solutions of the Einstein field 
equations have been obtained by assuming two different volumetric expansion laws in a way to cover all possible 
expansions: namely power-law and exponential-law expansion. To my knowledge, the literature has hardly witnessed 
this sort of exact solutions for the anisotropic B-I space-time. So the derived DE models add one more feather 
to the literature. 
\end{itemize}
\begin{itemize}
\item In both cases (i) \& (ii), it is observed that, in early stage, the EoS parameter $\omega$ is positive i.e. 
the universe was matter dominated in early stage but in late time, the universe is evolving with negative values 
i.e. the present epoch (see, Figures 1 \& 2). Thus our DE models represent realistic models.
\end{itemize}
\begin{itemize}
\item In both cases the DE models present the dynamics of EoS parameter $\omega$ provided by Eqs. (\ref{eq35}) and 
(\ref{eq60}) respectively may accommodated with the acceptable range  $-1.67 < \omega < -0.62$ of SNe Ia data 
(Knop et al. \cite{ref35}). It is already observed and shown in previous sections that for different cosmic times, 
we obtain cosmological constant dominated universe, quintessence and phantom fluid dominated universe (Caldwell 
\cite{ref74}), representing the different phases of the universe through out the evolving process. Unlike Robertson-
Walker (RW) metric Bianchi type metrics can admit a DE that wields an anisotropic EoS parameter according to the 
characteristics. The cosmological data - from the large-scale structures \cite{ref80} and Type Ia supernovae 
\cite{ref78,ref81} observations - do not rule out the possibility of an anisotropic DE either \cite{ref82,ref83}. 
Therefore, one can not rule out the possibility of anisotropic nature of DE at least in the frame-work of B-I 
space-time.
\end{itemize}
\begin{itemize}
\item Our DE models are of interest because in both cases the nature of decaying vacuum energy density $\Lambda(t)$ 
are supported by recent cosmological observations (Garnavich et al. \cite{ref1,ref2}; Perlmutter et al. 
\cite{ref3}$-$ \cite{ref5}; Riess et al. \cite{ref6,ref77,ref78}; Schmidt et al. \cite{ref7}). These observations 
on magnitude and red-shift of type Ia supernova suggest that our universe may be an accelerating one with induced 
cosmological density through the cosmological $\Lambda$-term. 
\end{itemize}
\begin{itemize}
\item
Though there are many suspects (candidates) such as cosmological constant, vacuum energy, scalar field, brane world, 
cosmological nuclear-energy, etc. as reported in the vast literature for DE, the proposed model in this paper at 
least presents a new candidate (EoS parameter) as a possible suspect for the DE.   
\end{itemize}
\section*{Acknowledgments} 
The first author (A. Pradhan) would like to thank the Laboratory of Information Technologies, Joint Institute for 
Nuclear Research, Dubna, Russia for providing facility and support where part of this work was carried out. 

\end{document}